\documentstyle[aps,prb,epsf,multicol]{revtex}
\begin{document}

\draft

\title{Indications of Unconventional Superconductivity  \\
     in Doped and Undoped Triangular Antiferromagnets}
\author{Matthias Vojta$^{(a,b)}$ and Elbio Dagotto$^{(a)}$}
\address{
  (a) Department of Physics, National High Magnetic Field Laboratory, 
      Florida State University,
      Tallahassee, FL 32306, USA \\
  (b) Institut f\"{u}r Theoretische Physik,
      Technische Universit\"{a}t Dresden, D-01062 Dresden, Germany
}
\date{Submitted for publication on May 7, 1998}
\maketitle

\begin{abstract}
The possibility of superconductivity in doped and undoped triangular antiferromagnets 
is discussed. 
Using the Bethe-Salpeter (B-S) equation, it is shown that the exchange of
RPA paramagnons on a triangular lattice Hubbard model 
leads to strong pairing correlations at and near half-filling.
The dominant states for this system
correspond to $d$-wave singlet (even-frequency) and
$s$-wave triplet (odd-frequency) pairing.
Analytical techniques applied to the hole-doped \mbox{$t$-$J$} model yield similar results. 
A \mbox{$t_1-t_2$} Hubbard model interpolating between square 
($t_1=0$) and triangular ($t_1=t_2$) lattices
has a tendency to only $d$-wave singlet pairing for $t_1/t_2 \leq 0.8$.
Experimental consequences for organic compounds $\kappa$-(BEDT-TTF)$_2$X are discussed.
\end{abstract}

\widetext
\begin{multicols}{2}
\narrowtext



Since the discovery of high-temperature superconductivity,
two-dimensional (2D) doped antiferromagnets (AF) 
have been studied intensively. Experimental results
and a large body of theoretical work suggest that the pairing state
is highly anisotropic and probably a $d_{x^2-y^2}$-singlet.
In addition, materials with
2D spin arrangements on non-square lattices have also been synthetized.
For example, experiments suggest the realization of a
triangular spin-$1 \over 2$ AF in NaTiO$_2$~\cite{Takane94},
as well as in surface structures~\cite{Surfaces} such as 
K/Si(111):B. Delafossite cuprates RCuO$_{2 + \delta}$,
with R a rare-earth element, have Cu ions sitting on a triangular
lattice~\cite{defa}. Other compounds with this geometry have also been
prepared~\cite{other}.
Recently, theoretical arguments in the context of organic
superconductors suggest that $\kappa$-(BEDT-TTF)$_2$X, where X is an ion,
is described by the half-filled
Hubbard model on an anisotropic triangular lattice with one electron 
per site~\cite{BEDTTTF}.
Metallicity is obtained by working below the critical coupling that leads
to an insulating phase.

The presence of unconventional superconductivity on a square-lattice
AF suggests that similar unusual phenomena may occur
in other 2D AF arrangements, such as a triangular lattice. 
Although most of the materials described above do not contain a 
finite concentration of mobile carriers, it is important to analyze on
theoretical grounds
the properties of holes in this environment and specially
the characteristics of a possible superconducting state. 
This information may induce further experimental work in non-square lattice doped spin
systems. 
Precisely in this paper the issue of pairing in triangular and anisotropic triangular AF 
is addressed.



The model to be discussed in the first part of this paper
contains a nearest-neighbor hopping term $t_2$ on a 2D square lattice 
and an additional hopping term $t_1$ along $one$ diagonal of each plaquette,
so that $t_1=0$ corresponds to a square lattice (bandwidth $8\,t_2$) whereas
$t_1=t_2$ is an isotropic triangular lattice (bandwidth $9\,t_2$).
For details see Ref.~\cite{BEDTTTF}.
The interaction will be modeled by an on-site 
Hubbard repulsion $U n_{i\uparrow} n_{i\downarrow}$. 

To study the effective interaction between electrons arising
from the exchange of paramagnons the RPA approximation is used, as done
before for the analogous problem on the square lattice~\cite{Paramagnons}
where further work~\cite{Review} using more sophisticated techniques showed that 
indeed indications of $d$-wave pairing are present in this context.
Then, the B-S approach is powerful enough to capture the essence of the
pairing tendencies in Hubbard-like models.
The effective pairing potentials for the singlet and
the triplet channels are
\begin{eqnarray}
V_s(p,p') &=& U
 + { U^3 \chi_0^2(p'-p) \over 1-U^2 \chi_0^2(p'-p)}
 + { U^2 \chi_0(p'+p) \over 1-U \chi_0(p'+p)},  \nonumber \\
V_t(p,p') &=& 
 - { U^2 \chi_0(p'-p) \over 1-U^2 \chi_0^2(p'-p)},
\end{eqnarray}
with $p=({\bf p},i\omega_n)$ and $\omega_n$ being a fermionic Matsubara 
frequency.
$\chi_0({\bf q},\omega)$ denotes the spin susceptibility:
\begin{equation}
\chi_0({\bf q},\omega) = {1 \over N} \sum_{\bf p}
{ f(\epsilon_{\bf p+q})-f(\epsilon_{\bf p}) \over
  \omega-(\epsilon_{\bf p+q}-\epsilon_{\bf p})+i0^+ },
\end{equation}
where $\epsilon_{\bf p}$ are the non-interacting electron energies.

\begin{figure}
\epsfxsize=7.8 truecm
\centerline{\epsffile{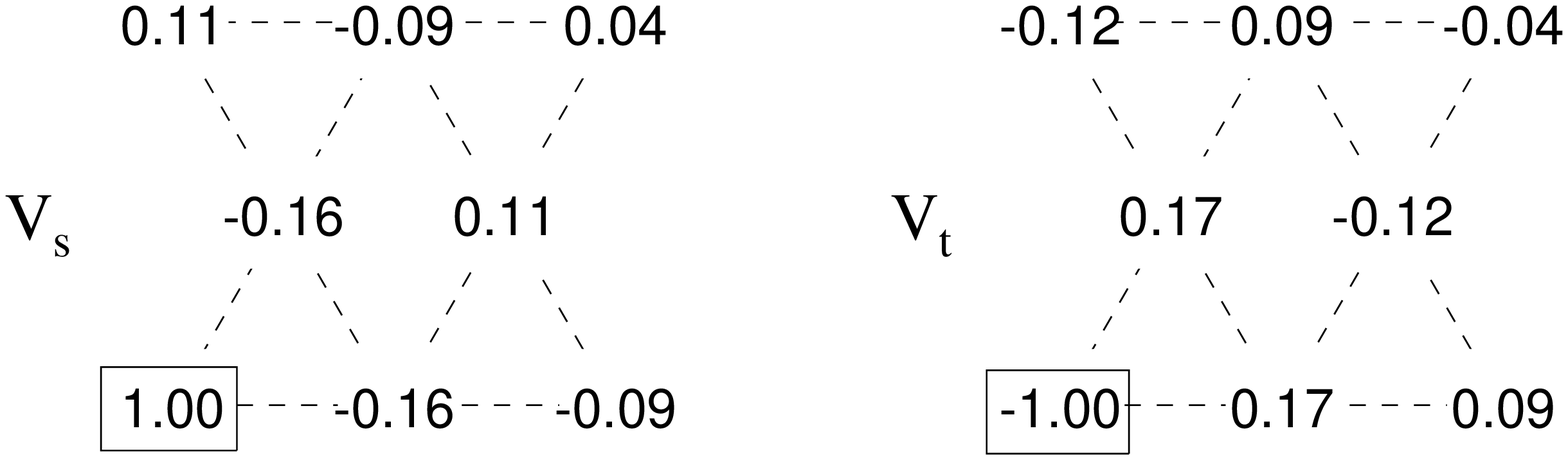}}
\caption{Effective singlet ($V_s$) and triplet ($V_t$) interaction in real-space 
for $U/t_2=4$, $t_1=t_2$, $\omega=0$ and a
density of $\langle n\rangle \approx 0.9$ 
(results at half-filling are similar). 
The numbers are normalized such that the on-site interaction has unit magnitude.
(The reference site is marked by a rectangle.) 
The ratio of on-site interactions is $V_s(0)/V_t(0)\approx -3$.
With decreasing band filling the range and magnitude of the interactions
decrease.}
\label{FIG_VEFF}
\end{figure}

These interactions Fourier-transformed into real-space at $t_1=t_2$ (triangular lattice)
and in the static limit ($\omega=0$)
are shown in Fig. \ref{FIG_VEFF}. The density is chosen to be slightly below
half filling, i.e., $\langle n\rangle \approx 0.9$.
The singlet pairing interaction is repulsive 
on-site, but attractive for pairs of electrons at nearest-neighbor
sites.
The triplet interaction $V_t$ behaves qualitatively similar to $V_s$
except for the difference in sign, i.e., it is repulsive for nearest-neighbor pairs. 
Note that the triangular system does not obey perfect nesting at 
half-filling.
Thus, the tendency to form a spin-density wave state is not as strong as on
the square lattice Hubbard model.
However, for large $U$ and half-filling the model maps onto the $S=\frac{1}{2}$
triangular AF~\cite{Fazekas74}
which is believed to have an ordered 120$^\circ$ magnetic ground state~\cite{THAF_LRO}.

\begin{figure}
\epsfxsize=8.2 truecm
\centerline{\epsffile{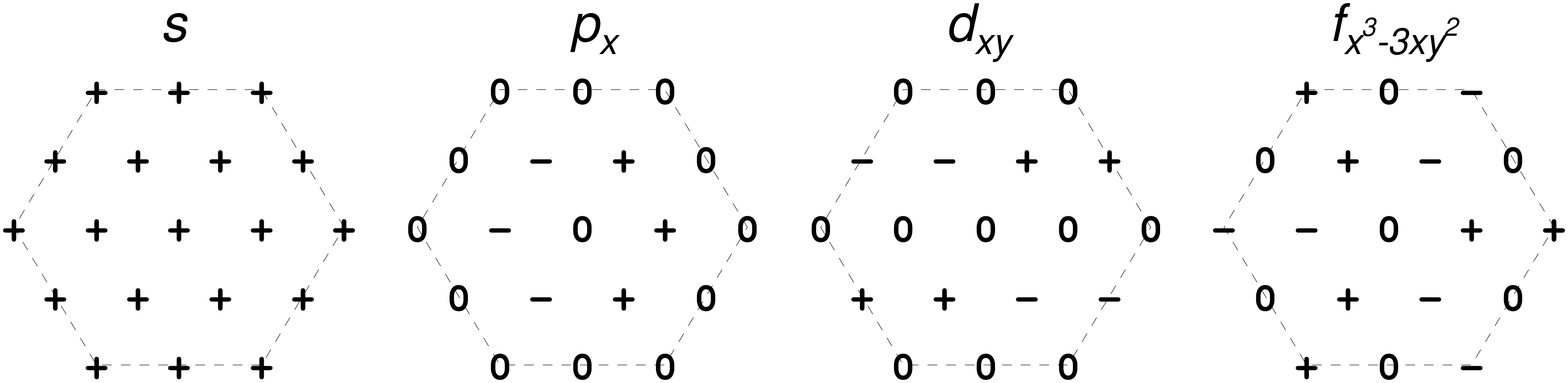}}
\caption{Momentum dependence of the pair wavefunctions corresponding
to $s$, $p_x$, $d_{xy}$, and $f_{x^3-3xy^2}$ symmetries.
The graphs show the sign and the zeros of the pair wavefunction at selected
points of the BZ.}
\label{FIG_CHANNELS}
\end{figure}

To investigate two-particle bound states it is convenient to classify them
according to symmetry properties in momentum space.
The point symmetry group of the isotropic triangular lattice is $C_6$ which has 
6 irreducible representations
($s$, $p_x$, $p_y$, \mbox{$d_{x^2-y^2}$,} \mbox{$d_{xy}$,} $f_{x^3-3xy^2}$). 
The results show that the triangular system has possible instabilities in the 
$s$-, $p_x$-, $d_{xy}$- and $f_{x^3-3xy^2}$-channels 
illustrated in Fig. \ref{FIG_CHANNELS}
using the Brillouin zone (BZ) of the triangular system.
There are 3 equivalent $d_{xy}$-solutions (related by 60$^\circ$ rotation),
one of them continuously evolves into the well-known 
square-lattice $d_{x^2-y^2}$-symmetry when reducing $t_1$ and going to the 
square BZ.

To search for possible pairing instabilities in the problem
let us here
investigate the leading eigenvalues $\lambda$ and
eigenfunctions $\phi(p)$ of the Bethe-Salpeter (B-S) equation
\begin{eqnarray}
\lambda \, \phi(p) = - {T \over N} \sum_{p'}
  V(p,p') G(p') G(-p') \phi(p')
\,.
\label{BETSAL}
\end{eqnarray}
Self-energy contributions are neglected, and the single-particle Green's function is
$G(p) = (i\omega_n - \epsilon_{\bf p})^{-1}$.
When the largest eigenvalue $\lambda$ reaches 1, a superconducting
instability to a state with a pair wavefunction $\phi(p)$ occurs. 

The B-S equation can have singlet and triplet solutions which correspond
to pair wavefunctions with even or odd parity when transforming from $({\bf p},i\omega_n)$
to $(-{\bf p},-i\omega_n)$. 
The usual $s$- and $d$-wave singlet solutions are even both in frequency and momentum
whereas the usual triplet solutions ($p$ and $f$) are even in frequency and odd in
momentum.
As discussed by Berenzinskii and by Balatsky and Abrahams \cite{OddOmega} also odd-frequency
pair wavefunctions are permitted by symmetry. 
Then, one can obtain even-momentum triplets 
($\phi(-{\bf p},i\omega_n) = \phi({\bf p},i\omega_n) =
  - \phi({\bf p},-i\omega_n) $)
and odd-momentum singlets
($\phi(-{\bf p},i\omega_n) = -\phi({\bf p},i\omega_n) =
  \phi({\bf p},-i\omega_n) $).
In a model with strong on-site repulsion there are two ways for the
pairs to "avoid" the repulsion. One is by constructing 
a pair wavefunction with zero on-site amplitude, i.e. with
non-zero angular momentum. The second way is by having
an odd-$\omega$ dependence which implies a
vanishing equal-time pair amplitude (and spontaneous breaking of time reversal
symmetry).

The present investigation shows that the dominant pairing channels 
in the triangular lattice model treated with the B-S equation 
are $d$-wave singlet ($even$ in frequency) at intermediate $U/t_2$ switching
to $s$-wave triplet ($odd$ in frequency) at large $U/t_2$,
although a 
small deviation from this geometry using the $t_1-t_2$ model establishes
a $d$-wave state as the only dominant one.
(For small $U/t_2 \leq 2.5$ and $t_1 = t_2$ the transition temperature is smaller
than $10^{-3}\,t_2$.)
This curious competition for the triangular lattice
can be understood intuitively as follows, starting with the
even-$\omega$ case first: 
The singlet potential favors pairing with zero on-site amplitude due to
the Hubbard repulsion. This effect suppresses $s$-wave singlets,
as is the case in the square lattice near
half-filling\cite{Paramagnons}.
In the family of even-$\omega$ solutions the next possibility is
a $d$-wave singlet. However, in contrast to the square lattice where 
$d_{x^2-y^2}$-pairing is favored\cite{Paramagnons},
the tendency towards $d$-wave solutions is not as strong in the triangular 
system.
This can be understood from the B-S equation: The susceptibility $\chi_0$ and the effective 
potential are strongly peaked at the ordering wavevector 
${\bf Q} = ({4 \over 3}\pi,0)$ for the triangular system. 
The kernel of the B-S equation for $T\rightarrow 0$ basically couples
values of $\phi(p)$ at
Fermi surface momenta which differ by $\bf Q$. 
The particular value of $\bf Q$ for the triangular system 
weakens $d_{x^2-y^2}$- and $d_{xy}$-wave singlet solutions.
How about $p$ and $f$ symmetries in the even-$\omega$ sector? Here
the effective triplet potential is repulsive for nearest neighbors,
and it tends to suppress
all triplet pairing with zero amplitude on-site.
This reasoning lead us to believe that odd-$\omega$ pairing may be relevant
for the triangular lattice. In the triplet channel,
the $s$-wave solution can take advantage of the on-site interaction
which remains attractive at non-equal times, thus
avoiding the suppression by the Pauli principle. 
Note that in previous B-S studies on the square lattice it was observed that
odd-$\omega$ solutions actually compete in strength with the even-$\omega$ $d$-wave
possibility~\cite{bulut}. 

\begin{figure}
\epsfxsize=6 truecm
\centerline{\epsffile{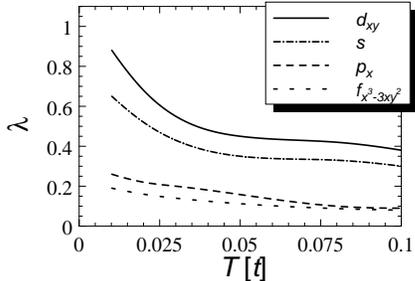}}
\caption{
  Bethe-Salpeter eigenvalues vs temperature for the isotropic
  triangular lattice, $U=3.5\,t_2$, and half-filling
  ($d$-wave singlets solid, $s$-wave triplets dash-dot, $p$-wave singlets dashed,  
  $f$-wave singlets dotted).
}
\label{FIG_BETSALHF}
\end{figure}

\begin{figure}
\epsfxsize=7.5 truecm
\centerline{\epsffile{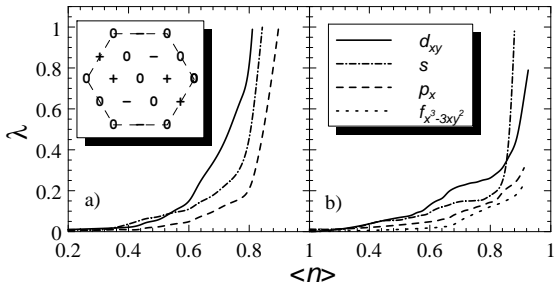}}
\caption{Bethe-Salpeter eigenvalues vs. band filling for $U=4\,t_2$.
a) $t_1/t_2=0.8$ and $T=0.08\,t_2$.
b) $t_1/t_2=1.0$ (triangular lattice) and $T=0.02\,t_2$. 
The inset in a) shows the symmetry of the $d$-wave solution in the triangular
lattice BZ which becomes $d_{x^2-y^2}$ for the square lattice case.
}
\label{FIG_BETSAL1}
\end{figure}

To verify this intuitive picture, the B-S equation was solved on a finite lattice 
(up to $32^2$ sites) with a cutoff on the Matsubara frequencies 
of $8\,t_2$.
The eigenvalues reach unity only at and close to half-filling. 
The transition temperature is determined by the temperature where this
largest eigenvalue reaches unity.
In Fig. \ref{FIG_BETSALHF} the leading eigenvalues are shown for the triangular lattice 
($t_1=t_2$) at half-filling, $U/t=3.5$, vs temperature. In this regime
the dominant eigenvalue corresponds to $d$-wave singlet pairing.
Small deviations from the perfect triangular lattice using
the $t_1-t_2$ model also give a stable $d$-wave singlet solution 
at a temperature \mbox{$T=0.08\,t_2$}
(Fig. \ref{FIG_BETSAL1}a).
This regime is likely of relevance for organic
superconductors~\cite{BEDTTTF}.
However, it is interesting to observe that for $U/t \geq 4$ and half-filling
the $d$-wave even-$\omega$ and the three odd-$\omega$ solutions strongly 
compete in the considered temperature range \mbox{$0.01 < T/t_2 < 0.1$}
(Fig. \ref{FIG_BETSAL1}b).
Actually, the eigenvalue corresponding to the $s$-wave pairing instability
is the largest in this regime, and thus odd-$\omega$ $s$-wave triplets
are favored, as expected from the previous discussion.
Since self-energy contributions are neglected 
the critical temperature for the pairing instability, which is around
$0.02\,t_2$, should only be considered as a rough estimate. Compared with
the results obtained on a square lattice~\cite{bulut} ($\sim 0.1\,t_2$), the
triangular lattice seems to have a critical temperature substantially
smaller. 

Thus far, the calculations have addressed the weak coupling limit.
In the second part of this paper an isotropic triangular AF
doped with holes described by the $t$-$J$ model will be discussed.
At half-filling the \mbox{$t$-$J$} model reduces to the triangular Heisenberg
antiferromagnet (THAF)\cite{Fazekas74} with spins $1 \over 2$.
It is by now widely believed
that its ground state possesses magnetic long-range order
\cite{THAF_LRO} which can be described as 120$^\circ$-order,
namely the ground state of the classical spin model with the addition of quantum 
fluctuations.

\begin{figure}
\epsfxsize=8.3 truecm
\centerline{\epsffile{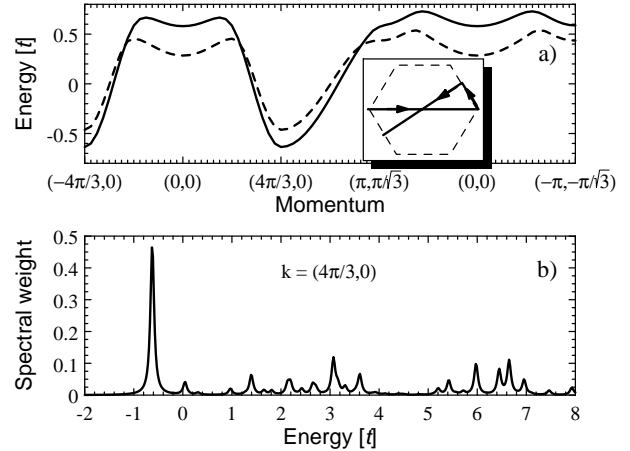}}
\caption{ a) One-hole dispersion calculated from the polaron picture at
  $J/t = 0.1$ (dashed) and 0.4 (solid). The energy zero level has been set
  at the center of mass of the band. 
  The inset shows the path in momentum space corresponding to the horizontal 
  axis.
  b) One-hole spectral function at $J/t=0.4$ and momentum $({4 \over 3}\pi,0)$
  which is the minimum of the quasiparticle (QP) band. A sharp QP
  peak is observed.}
\label{FIG_ONEHOLE}
\end{figure}

The one-hole problem in the triangular $t$-$J$ model was studied
analytically before
using the self-consistent Born 
approximation (SCBA) \cite{Azzouz,Apel}.
Since the SCBA neglects closed loops in the hole motion 
(which are more important here than for the square
lattice since only three hopping steps are necessary for one loop)
in this paper an alternative analytical approach based on the 
picture of magnetic polarons will be used.
The spin excitations are described by a set of string operators
locally distorting the background spin configuration
\cite{Strings,Riera97,Vojta98}
which is assumed to be long-range ordered.
The one-hole spectral function can be evaluated using the
Mori-Zwanzig projection technique. 
%
For details see Refs.~\cite{Vojta98,Vojta99}.
Fig. \ref{FIG_ONEHOLE}a contains the calculated one-hole dispersion $\epsilon_{\bf p}^{\rm hole}$
for different couplings $J/t$.
In Fig. \ref{FIG_ONEHOLE}b the one-hole spectral function at
$J/t=0.4$ and the momentum corresponding to the band minimum 
${\bf p} = ({4 \over 3}\pi,0)$ is shown. A sharp peak at the bottom of
the spectrum suggests that a description in terms of quasiparticles (QP) is
appropriate. 
Note that the one-hole dispersion shown in Fig. \ref{FIG_ONEHOLE}a is in 
qualitative agreement with exact diagonalization (ED) data on a 12-site cluster~\cite{Azzouz}
as well as with SCBA and 21-site ED results from Ref.~\cite{Apel}.
(A detailed discussion will be given elsewhere~\cite{Vojta99}.)

Now let us consider a fermionic model
constructed from the dressed one-hole properties and including an
interaction between the QP. 
The use of a dressed QP dispersion implicitly assumes that in the normal state 
short-range antiferromagnetism exists, namely the AF correlation length is not 
negligible. This is a standard assumption in AF scenarios for the cuprates
and can be checked experimentally using, e.g., NMR techniques.
For simplicity, the QP interaction is taken from the RPA calculation described in the
first part of the paper.
For singlet pairing it basically consists of an on-site repulsion,
a nearest-neighbor attraction, and a NNN repulsion.
Note here that the $\omega$-dependence of the interaction is important
for possible odd-$\omega$ pairing.
Previous literature in the context of the square lattice ~\cite{Review,Nazarenko} 
has shown that in $both$ weak and strong coupling the most important part of
the interaction for the study of pairing is the NN attraction (singlet channel).
Thus, for the qualitative purposes of the present strong coupling study it is reasonable
to assume that a similar result holds in triangular lattices.

\begin{figure}
\epsfxsize=6 truecm
\centerline{\epsffile{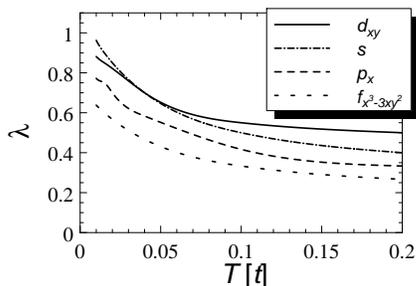}}
\caption{Temperature dependence of the Bethe-Salpeter eigenvalues 
for $s$-wave triplets, and $p$-wave, $d$-wave singlets, 
and $f$-wave singlets, now
calculated from the strong-coupling dispersion and the RPA interaction.}
\label{FIG_BETSAL2}
\end{figure}

To investigate the existence of pairs within this model once again
the B-S equation Eq.(\ref{BETSAL}) will be solved. Now
the single-particle Green's function is determined by the one-hole
QP dispersion from the strong-coupling calculation:
$G(p) = (i\omega_n - \epsilon_{\bf p}^{\rm hole})^{-1}$.
For the calculation the RPA interaction $V(p,p')$ corresponding
to an electronic density $\langle n\rangle = 0.9$ will be used.
In Fig. \ref{FIG_BETSAL2} the leading eigenvalues are plotted 
as a function of temperature for 
the different pairing symmetries, working with a
low-band $hole$ filling (10\%).
The results qualitatively 
agree with those obtained in  the Hubbard model.

Odd-frequency pairing such as the one
described here has experimental interesting
consequences, as discussed in Ref.~\cite{OddProperties}.
First, there is no gap in the excitation spectrum at any $\bf k$.
Second, the magnitude of the Hebel-Slichter peak in the NMR relaxation rate
is strongly reduced compared to the BCS case.
Other nontrivial consequences of odd-frequency pairing include
the tunneling between an odd-frequency and a BCS 
superconductor. 

However, 
regarding the organic superconductors $\kappa$-(BEDT-TTF)$_2$X, our results
suggest that their low-temperature properties will likely correspond
to a $d$-wave
singlet state being the analog of the widely discussed $d_{x^2-y^2}$-state of
square lattices which arises from antiferromagnetic fluctuations.



Summarizing, in this paper the possibility of superconductivity in
doped and undoped triangular and anisotropic antiferromagnets has been 
investigated.
The pairing interaction is assumed to arise from the exchange of
paramagnons. Both 
weak-coupling and strong-coupling approaches suggest that the system
presents a pairing instability in an unconventional channel.
A $d$-wave singlet superconducting state similar to the square lattice
dominates for anisotropic lattices, and even for the isotropic ones at
intermediate $U/t$. However, at large $U/t$ pairing in the
$s$-wave triplet odd-frequency channel is also possible.
Hopefully the present calculations will motivate experimentalists to
study in more detail antiferromagnetic triangular lattice compounds. 
Recent results~\cite{BEDTTTF} suggest that the physics described here
may already be realized in organic superconductors. 
If ARPES experiments were possible in this context~\cite{fujimori}, 
nodes should be observed for a $d$-wave singlet state.
Specific heat and 
penetration depth measurements \cite{BEDTTTF,SpecHeat} already suggest 
the existence of gapless excitations in these materials.


The authors thank J. A. Riera, R. H. McKenzie, and A. V. Dotsenko for 
useful conversations.
M.V. acknowledges support by the DAAD (D/96/34050) and the hospitality of the NHMFL
where part of this work was performed. 
E. D. is supported by the NSF grant DMR-9520776.



\end{multicols}

\end{document}